\documentclass[acmtog]{acmart}

\usepackage{booktabs} 

\usepackage{soul}
\usepackage{multirow}
\usepackage{multicol}
\usepackage{subcaption}
\usepackage{array}
\usepackage{hyperref}
\usepackage{wrapfig}

\newcolumntype{C}[1]{>{\centering\arraybackslash}p{#1}}

\usepackage[ruled]{algorithm2e} 

\SetAlFnt{\small}
\SetAlCapFnt{\small}
\SetAlCapNameFnt{\small}
\SetAlCapHSkip{0pt}

\acmJournal{TOG}

\begin{document}
\title{Neural ShDF: Reviving an Efficient and Consistent Mesh Segmentation Method}

\author{Bruno Roy}
\email{bruno.roy@autodesk.com}
\affiliation{
  \institution{Autodesk Research}
  \city{Montreal}
  \country{Canada}
}

\begin{abstract}
Partitioning a polygonal mesh into meaningful parts can be challenging. Many applications require decomposing such structures for further processing in computer graphics. In the last decade, several methods were proposed to tackle this problem, at the cost of intensive computational times. Recently, machine learning has proven to be effective for the segmentation task on 3D structures. Nevertheless, these state-of-the-art methods are often hardly generalizable and require dividing the learned model into several specific classes of objects to avoid overfitting. We present a data-driven approach leveraging deep learning to encode a mapping function prior to mesh segmentation for multiple applications. Our network reproduces a neighborhood map using our knowledge of the \textsl{Shape Diameter Function} (ShDF) method using similarities among vertex neighborhoods. Our approach is resolution-agnostic as we downsample the input meshes and query the full-resolution structure solely for neighborhood contributions. Using our predicted ShDF values, we can inject the resulting structure into a graph-cut algorithm to generate an efficient and robust mesh segmentation while considerably reducing the required computation times.
\end{abstract}

%
%



\maketitle


\begin{figure}[t]
    \centering
    \includegraphics[width=0.85\linewidth]{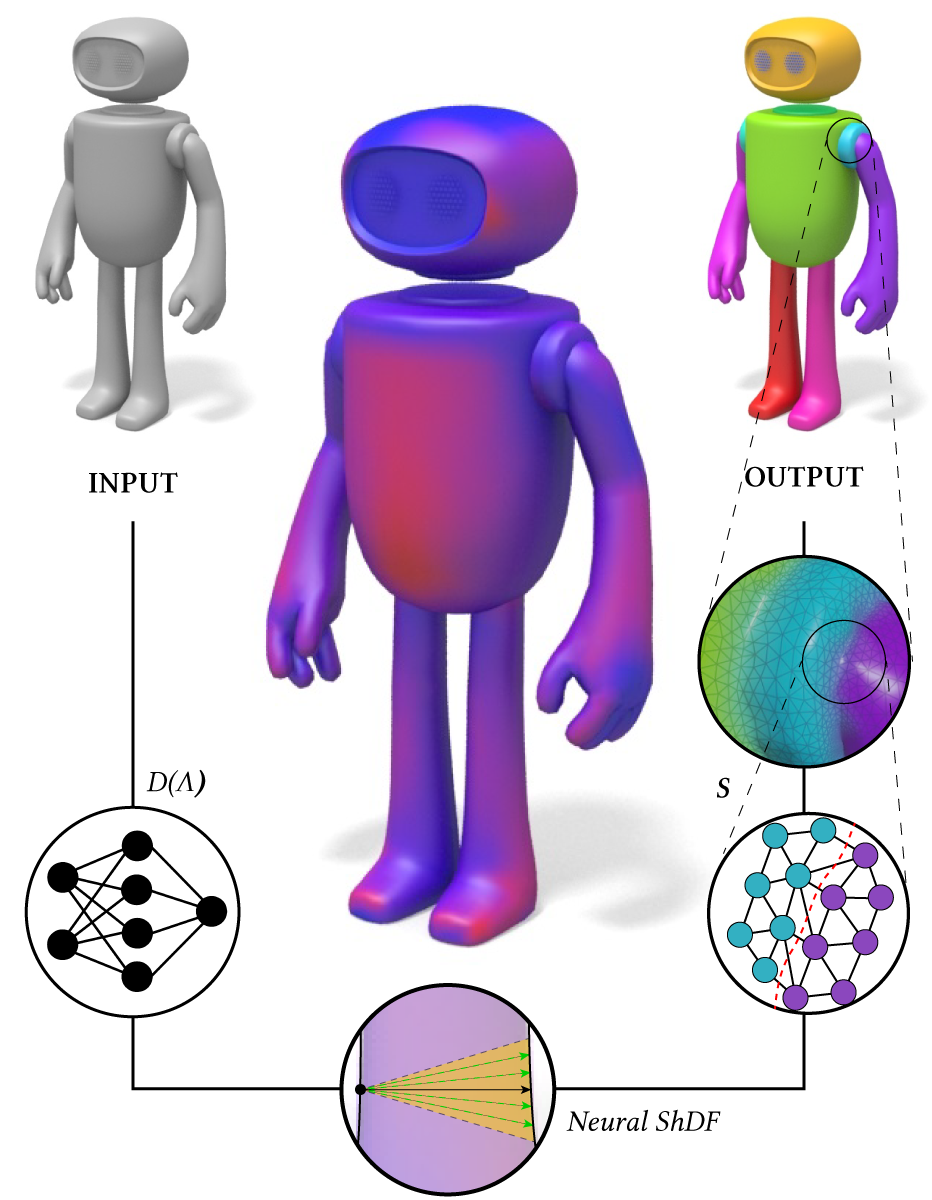}
    \caption{Our approach introduces a neural approximation of the \textit{Shape Diameter Function}. We use the predicted values of our network to provide a highly adaptive and controllable mesh segmentation workflow.}
    \label{fig:teaser}
\end{figure}

\section{Introduction}
Mesh segmentation is the process of decomposing a polygonal mesh into meaningful parts. These meaningful parts consist of subsets of vertices or faces that are semantically relevant and informative for further applications. Decomposing such structures is usually performed by defining edge loops (i.e., pairs of vertices) acting as boundaries between these subsets of elements. Although performing this task on a closed manifold mesh seems inherently intuitive, generalizing is still challenging considering that multiple valid solutions exist for any mesh. Moreover, most of the state-of-the-art methods are rather time-consuming.

The problem with current mesh segmentation methods is that they mostly rely on constraints -- making them highly time-consuming as they require iterative solvers to jointly optimize them. The constraint types are generally cardinality, geometry, and topological, respectively guiding the number of segmented parts, biases towards specific primitive shapes, and sub-mesh connected components. Some of these methods also require the user to select initial seeds to reduce the number of iterations before satisfying the constraints, adding on the total time required to accomplish the mesh segmentation task.

In recent years, machine learning has revived the problem of segmentation on meshes by proposing much more generalizable approaches and by offering a better understanding of the intrinsic semantics of shapes. Although these approaches have considerably eased this challenging task, the time required to train and cover a wide spectrum of different topologies remains problematic for use on dense meshes.

In this work, we propose a data-driven method leveraging a correspondence between the surface and its underlying volume to efficiently segment a polygonal mesh. We generate this correspondence by using a measure based on the medial axis transform from the \textit{Shape Diameter Function} (ShDF). The neural ShDF values are then used as inputs to a graph-cut algorithm providing an efficient and controllable workflow for mesh segmentation. By combining our Neural ShDF with a state-of-the-art graph-cut algorithm, our proposed method is capable of generating a high-quality segmentation for any polygonal mesh at a fraction of the cost. In addition, our approach provides an intuitive way to generate various solutions for mesh segmentation by reusing the neural ShDF values with different sets of parameters throughout the graph-cut steps. The key contributions are as follows:
\begin{itemize}
    \item{We introduce a neural \textit{Shape Diameter Function} improving generalization of local feature for mesh segmentation.}
    \item{We propose a novel approach reusing neural ShDF values to generate various and unique outcomes -- making it efficient and highly adaptive.}
    \item{We propose a resolution-agnostic approach by downsampling the input mesh while solely querying the full-resolution mesh for neighborhoods.}
    \item{We provide experimental results on various applications of mesh segmentation for computer graphics workflows.}
\end{itemize}

\begin{figure*}[t]
  \centering
   \includegraphics[width=\textwidth]{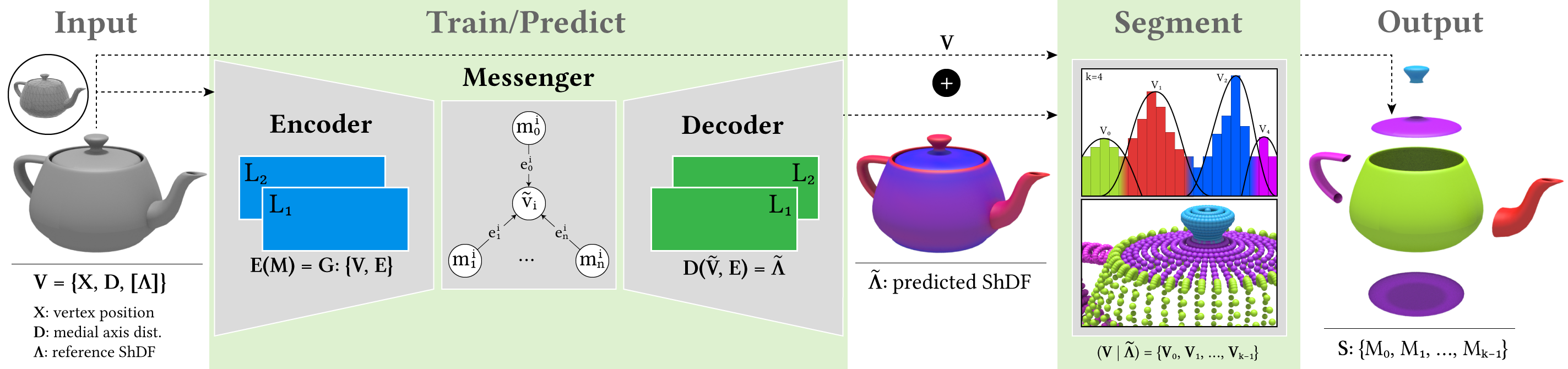}
   \caption{This is our \textit{Neural Shape Diameter} approach workflow. We train using both full-resolution (for neighbors when updating nodes by the \textbf{Messenger}) and coarse-resolution meshes (downsampled for training) with our Encoder-Messenger-Decoder ($\textbf{EMD}$) network. The $\textbf{EMD}$ network generates the predicted ShDF values $\tilde{\Lambda}$ used by our k-way graph-cut algorithm (i.e., gaussian mixture and clustering on GPU) to partition the final segmentation.}
   \label{fig:workflow}
\end{figure*}

\section{Related Work}
Over the last two decades, mesh segmentation has been used for various applications in computer graphics. This task has proven to benefit many applications in 3D shape analysis such as texture mapping~\cite{sander2003multi}, 3D shape modeling~\cite{ji2006easy}, 3D shape retrieval~\cite{ferreira2010thesaurus}, multi-resolution and mesh compression~\cite{maglo2011cluster}, and animation~\cite{yuan2016space}. We refer the reader to this survey for further details~\cite{rodrigues2018part} on part-based mesh segmentation. Historically, the traditional problem of mesh segmentation has been approached in many ways.
\paragraph{\textbf{Region growing.}}
One of the most intuitive and simple approaches for segmentation is the \textit{region growing} technique. The criterion that determines whether or not an element should be added to a cluster is what mainly differentiates the variations of the available \textit{region growing} algorithms. Among other criteria: representative planes~\cite{kalvin1996superfaces}, curvature~\cite{lavoue2005new}, and convexity~\cite{chazelle1995strategies, sheffer2007shuffler} were used as conditioners for clustering. Another common variation of the \textit{region growing} method using multiple source seeds to initiate the growing process~\cite{levy2002least, sorkine2002bounded, eck1995multiresolution}.
\paragraph{\textbf{Clustering.}} 
Merge operations on clusters is also crucial when it comes to segmentation tasks~\cite {garland2001hierarchical, attene2006hierarchical, gelfand2004shape, sander2001texture}. Although \textit{hierarchical clustering} algorithms are similar to \textit{growing regions} algorithms, these algorithms prioritize set operations between existing clusters in a structured way. \textit{Iterative clustering} algorithms are stated as parametric as the number of clusters is given a-priori. As opposed to previous methods, \textit{iterative clustering} methods are focusing on converging towards optimal segmentation given a number of clusters~\cite{lloyd1982least, hart2000pattern, shlafman2002metamorphosis, cohen2004variational, wu2005structure}.
\paragraph{\textbf{Implicit methods.}}
The \textit{implicit methods} for mesh segmentation are the closest in spirit to ours as they focus on boundaries and correspondences between subsets of elements of the object to segment. Again, the main difference between these algorithms is how they define the boundaries and underlying structures. To mention a few, the most common ways pass through curvature/contours~\cite{lee2005mesh, levy2002least, mitani2004making}, subdivision~\cite{katz2003hierarchical, podolak2006planar} (similar to hierarchical clustering), and underlying structures connecting the shape with intrinsic surface properties~\cite{li2001decomposing, raab2004virtual, lien2006simultaneous}. As a matter of fact, the \textit{Shape Diameter Function} algorithm~\cite{shapira2008consistent} is at the intersection of using subdivision and an underlying structure. The intuiting behind the \textit{Shape Diameter Function} is to produce a diameter measure throughout vertex neighborhoods of the mesh. The resulting measure (i.e., ShDF values) relates to the medial axis transform and provides a volume correspondence of the shape at any given point on the surface. These per-vertex measures are then used as a threshold for the graph-cut algorithm. In our approach, we take advantage of the generalization power of neural networks to estimate per-vertex properties as input to a graph-cut algorithm -- making the latter highly adaptive for mesh segmentation.
\paragraph{\textbf{Data-driven.}}
Neural networks have been widely used for the segmentation problem on images~\cite{lai2015deep}, point clouds~\cite{qi2017pointnet}, and more recently, meshes~\cite{hanocka2019meshcnn}. Several interesting approaches were proposed lately to tackle the mesh segmentation problem using deep neural network algorithms such as using convolution operators on edges~\cite{hanocka2019meshcnn}, converting 3D shapes into voxel-based representations~\cite{wu20153d, wang2017cnn, graham20183d}, and leveraging local features of point clouds~\cite{qi2017pointnet, qi2017pointnet++}. Nevertheless, these state-of-the-art data-driven methods remain hardly generalizable and often require dividing the learned model into several specific classes of objects to avoid overfitting. In contrast to these methods and Kovacicet et al.~\cite{kovacic2010fast}, we avoid the computationally expensive part by using predicted ShDF values $\tilde{\Lambda}$ to generate the final mesh segmentation. Moreover, learning a mapping function as opposed to directly learning to classify mesh elements~\cite{kalogerakis2010learning} (e.g., to cluster vertices) makes our approach more robust when used in unknown settings.

\section{Method}
As a formal definition, the traditional mesh segmentation task is described as follows: given a closed manifold mesh $\textbf{M}$, and $\textbf{E}$ the set of mesh elements (i.e., vertices, edges, or faces). The segmentation $\textbf{S}$ of mesh $\textbf{M}$ is a set of sub-meshes $\textbf{S}=\{M_0, ..., M_{n-1}\}$, where $M_i$ is defined as a subset of elements $e \in \textbf{E}$.

\subsection{Neural ShDF}
In the traditional \textit{Shape Diameter Function} method, the set of sub-meshes $\textbf{S}$ is obtained by subdividing a graph using the diameter measures $\lambda_i$ from per-vertex neighborhoods $\dot{v}_i$ as thresholds for the graph-cut algorithm. As we aim to estimate these ShDF values through a graph neural network $\textbf{EMD}$, the problem of finding the sub-meshes $M_i$ can be expressed as partitioning $\textbf{S}$ such that the constraint criteria $\{C_0, ..., C_n\}$ are minimized. The predicted ShDF values $\tilde{\lambda}_i$ are then used as threshold criteria to a graph-cut algorithm to generate the sub-meshes $\textbf{S}=\{M_0, ..., M_{n-1}\}$.

The constraint criteria $\{C_0, ..., C_n\}$ are defined as two terms: the similarities between the reference ShDF values $\lambda_i$ and the predicted ShDF values $\tilde{\lambda}_i$, and the neighborhood densities $\rho_i$ of vertices. We express the $L_2$ similarities between the reference ShDF values and the predicted ones as follows:
\begin{equation}
\label{eq:loss_SDF}
L_{\Lambda}=\frac{1}{n_j}\sum_{i}^{n_j}{\alpha\|\lambda_i-\tilde{\lambda}_i\|_2},
\end{equation}
where $n$ is the number of downsampled surface vertices (i.e., used to query the neighborhoods). Along with the input vertices, we provide to our network $\textbf{EMD}$ an additional term to weigh in the local mesh density for adaptive resolution meshes. We use the Poisson sampling disk method at the surface of meshes to select the downsampled vertices and compute their neighborhood density. The resulting densities $\rho_i$ are then used as a scaling factor during the messaging stage within our network to properly propagate the attributes in the current neighborhood.

\subsection{Model Architecture}
Our network $\textbf{EMD}$ architecture is essentially based on the Encode-Message-Decode model. The network is composed of two hidden MLP layers on both Encoder and Messenger with an output size of 128. The resulting Decoder output size matches the downsampled vertices provided as input during training and inference. Our model is trained and loaded for inference using an A6000 GPU with the Adam optimizer for 5M training steps with an exponential learning rate decay from $10^{-3}$ to $10^{-5}$ when passing the threshold of 3M steps.

\subsubsection{\textbf{Resolution-Agnostic Graph Network}}
\begin{wrapfigure}[7]{r}[0.25\columnwidth]{4.5cm}
\vspace*{-0.6cm}
\hspace*{-0.6cm}
\includegraphics[width=0.15\textwidth]{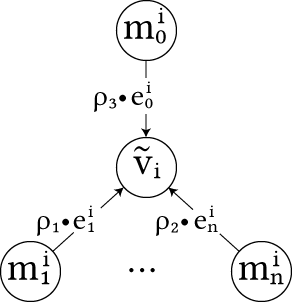} 
\label{fig:neighborhood}
\end{wrapfigure}
We handle varying resolutions using two mechanisms: downsampling the input mesh while maintaining full-resolution neighborhoods and using the neighborhood densities $\rho_i$ for each node $m^i$ as scaling factors when updating the current nodes during the messaging stage. The idea to remain resolution-agnostic is to pass the full-resolution neighborhoods along the downsampled mesh. That way, we can query the vertex neighborhoods by solely using the downsampled mesh vertices to compute the ShDF values $\tilde{\Lambda}$. The density values are used to scale the contributions of the neighbors through the edges $e^i$ during the message-passing step in our network. Moreover, as our network architecture requires a known input size, we use a fixed radius with the Poisson disk sampling algorithm to compute the neighborhood densities.

\subsubsection{\textbf{Graph Cut}}
\label{sec:graph_cut}
Once the predicted ShDF values $\tilde{\lambda}$ are obtained, we use them as inputs to a fast $k$-way graph-cut algorithm to offer an efficient and flexible way for mesh segmentation. Similarly to Fig.~\ref{fig:optimizing_params} in the appendices, our approach can leverage the predicted ShDF values using the grid-search method to find the optimal parameters for the mesh segmentation. We use a GPU implementation of the $k$-way graph partitioning as our graph-cut algorithm leveraging the \href{https://developer.nvidia.com/nvgraph}{nvGRAPH library} from NVIDIA.

Similarity to~\cite{shapira2008consistent}, our partitioning algorithm is composed of two steps. The first uses soft-clustering of the mesh elements (faces) to compute $k$ clusters based on their ShDF values, and the second finds the actual partitioning using $k$-way graph-cut to include local mesh geometric properties. Note that $k$, the number of clusters chosen, is more naturally related to the number of levels in the hierarchy and not to the number of parts.

\begin{figure}[h]
  \centering
   \includegraphics[width=\linewidth]{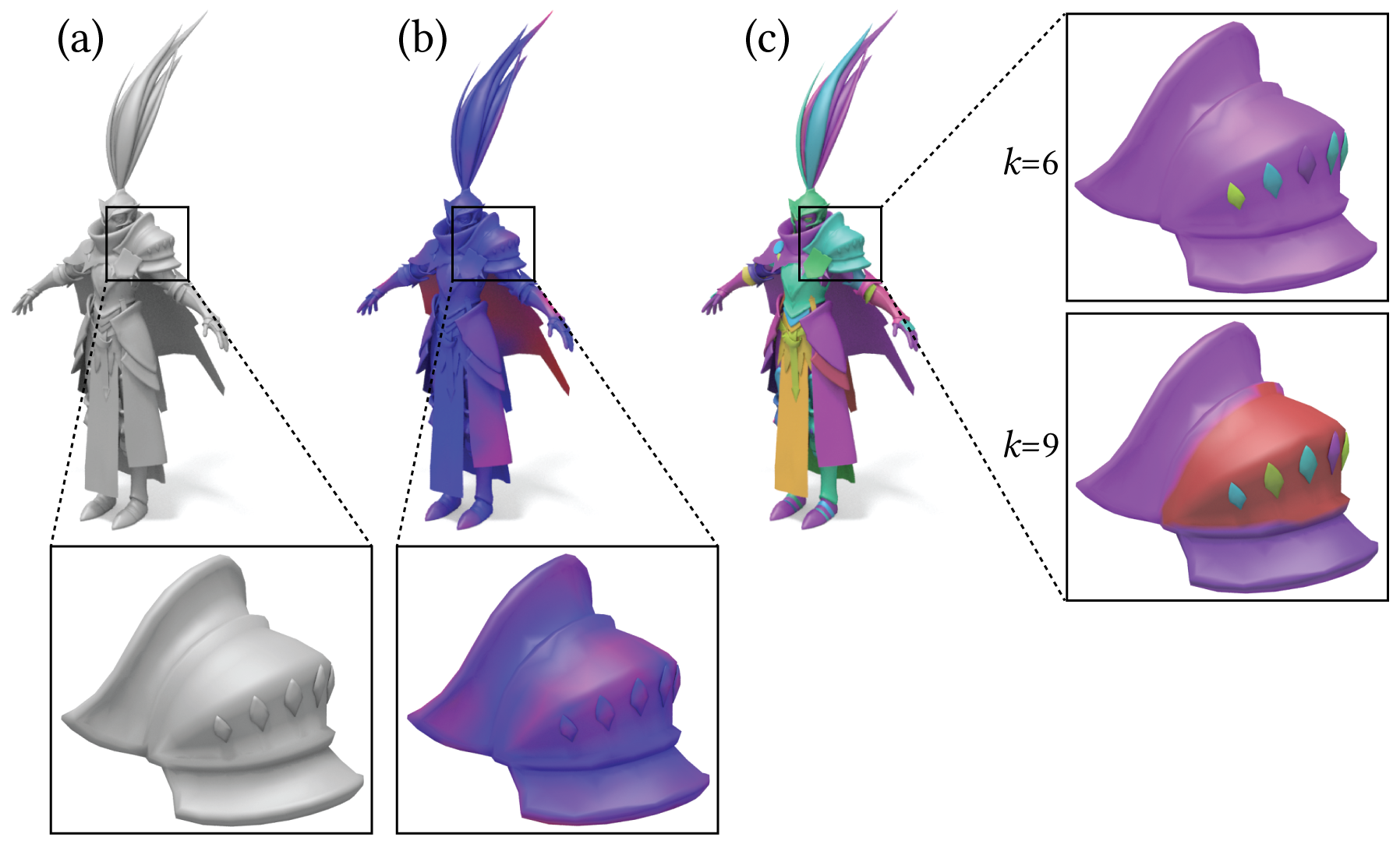}
   \caption{Example of the recursive refinement step on the \textsl{Samurai} model.}
   \label{fig:refinement}
\end{figure}

\begin{figure}[h]
  \centering
   \includegraphics[width=\linewidth]{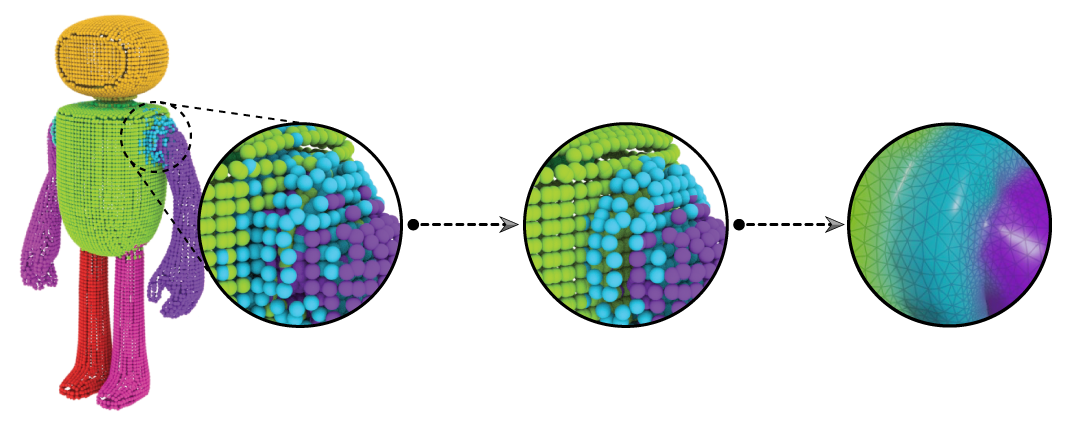}
   \caption{Post-processing leveraging the alpha expansion to smooth the segmentation boundaries.} 
   \label{fig:smooth_boundaries}
\end{figure}

As an optional post-processing step, we recursively use our approach as a refinement process to improve the segmentation of detailed parts (as shown in Fig.~\ref{fig:refinement}). In a complementary manner, we also performed on some of the presented results a post-processing step to reduce the noise close to the boundaries of the dense meshes. We update the boundaries between segmented parts by using an alpha expansion graph-cut algorithm considering both smoothness and concavity measures by looking at the dihedral angles between faces (as shown on the \textsl{Mayabot} in Fig.~\ref{fig:smooth_boundaries}).

\begin{figure}[t]
  \centering
   \includegraphics[width=\linewidth]{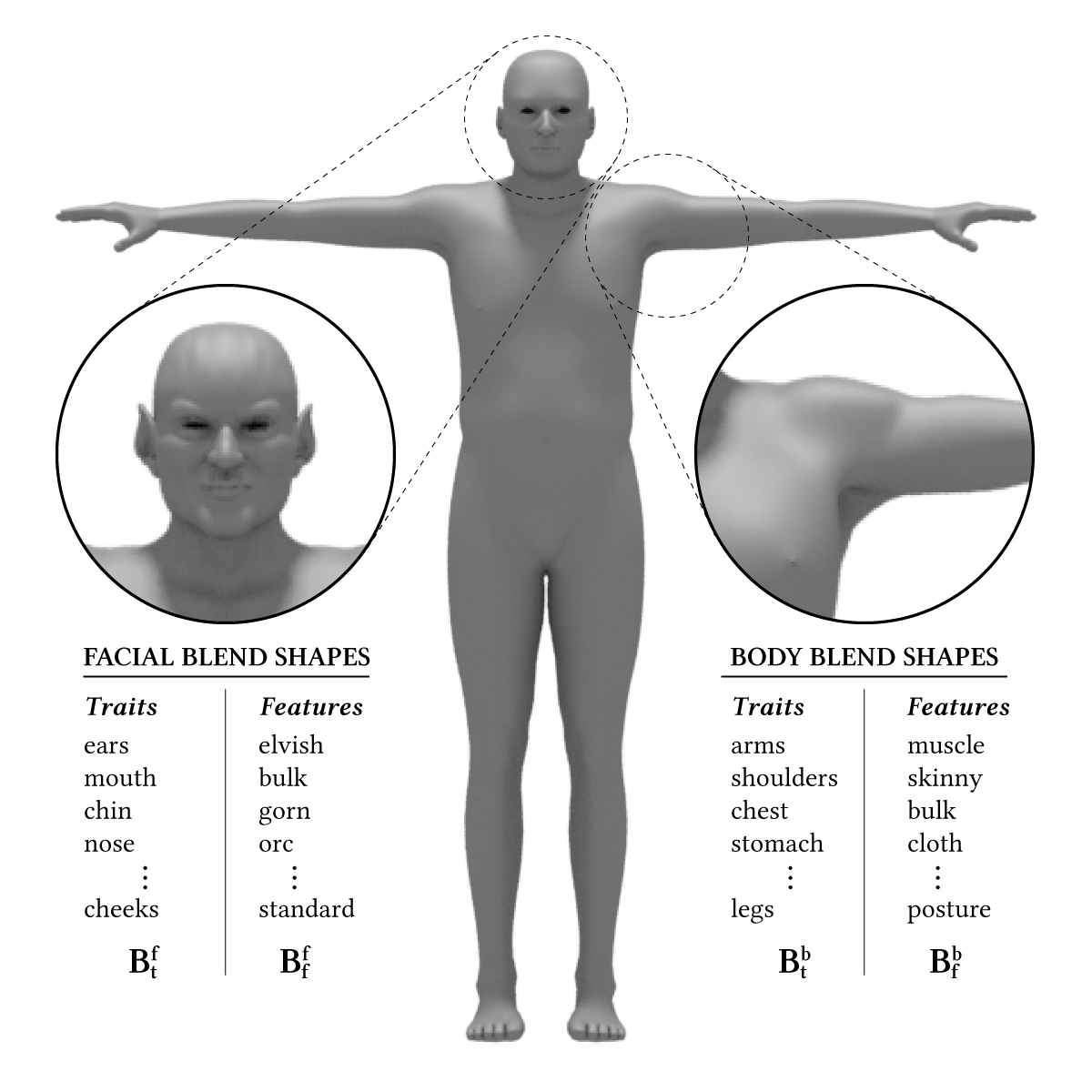}
   \caption{Base mesh and a subset of blend shapes used to generate the Autodesk Character Generator (ACG) dataset.} 
   \label{fig:dataset}
\end{figure}

\section{Dataset}
The dataset used to train our network $\textbf{EMD}$ is generated using the \textsl{Autodesk Character Generator}(ACG)\copyright~tool producing multiple variants of the same base mesh by applying blend shape operations on it. The generated dataset is self-consistent since it is exclusively composed of meshes having the same number of vertices. To train our network to encode resolution-agnostic features, we used a tessellation technique during training to provide multiple versions of the same mesh. We also used a re-meshing method to make our network resilient to consistent input samples (i.e., by changing the positions of selected downsampled vertices). We only use the tessellated and re-meshed meshes for neighborhood contributions during training.

\begin{figure}
    \centering
    \includegraphics[width=\linewidth]{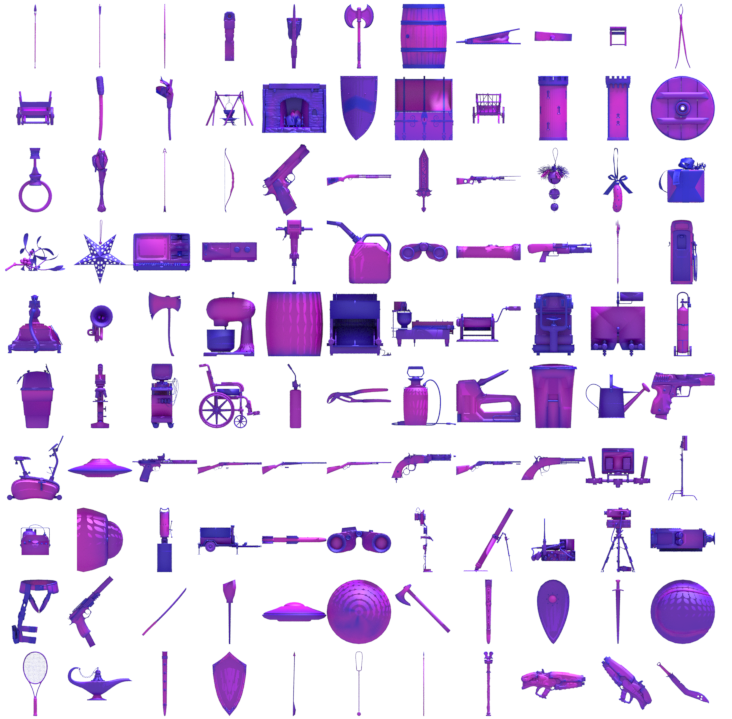}
    \caption{TurboSquid (TS) dataset: subset of assets.}
    \label{fig:subset_SDF}
\end{figure}

As shown in Fig.~\ref{fig:dataset}, we use several blend shapes $\textbf{B}$ to alter a base mesh into many variants to grow our dataset. The blend shapes $\textbf{B}$ are divided into two groups: facial $\textbf{B}^{f}$ and body $\textbf{B}^{b}$. For both groups, the blend shapes used to augment our dataset are described as either traits $\textbf{B}_{t}$ or features $\textbf{B}_{f}$. By permuting these blend shapes, we can generate a large dataset using solely a single base mesh. Additionally, we use an animation skeleton to randomly generate different poses as it may generate different ShDF values for the same mesh. Similarly to \cite{shapira2008consistent}, we use anisotropic smoothing on the ShDF values to overcome these differences during training. Moreover, we have built an additional custom dataset using TurboSquid (TS) assets (Fig.~\ref{fig:subset_SDF}) to evaluate our approach with production-ready content.

\section{Experiments and Results}
We evaluated our approach against a variety of state-of-the-art methods to highly its efficiency and precision while remaining controllable for the users. In the following sections, we compared our approach with a few baselines using known segmentation datasets and our own. We also present computation times on several scenarios using our approach as a few ones from previous work. We briefly discuss a particular case where our approach performs better than the state-of-the-art on dense meshes. Lastly, we will present a few applications that we believe would be useful to improve traditional graphics workflows.

\begin{figure}[h]
     \centering
     \begin{subfigure}{0.125\textwidth}
         \centering
         \includegraphics[width=\textwidth]{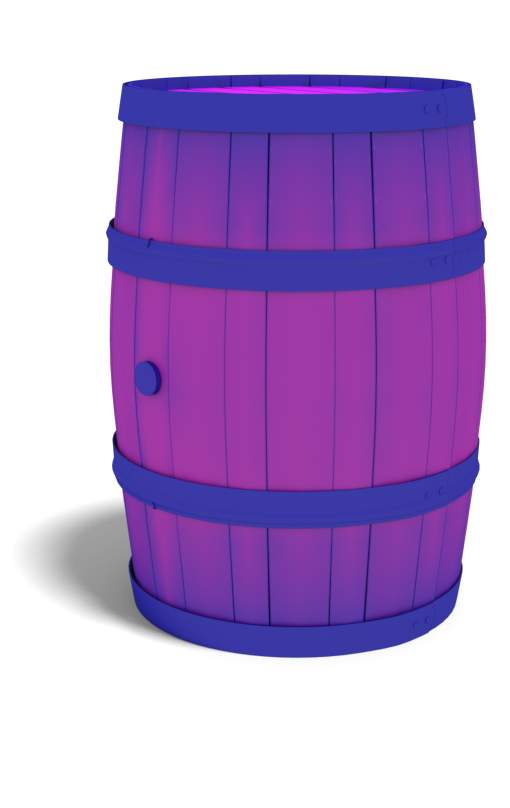}
         \caption{Reference}
         \label{fig:sdf_seg}
     \end{subfigure}
     \hfill
     \begin{subfigure}{0.2098\textwidth}
         \centering
         \includegraphics[width=\textwidth]{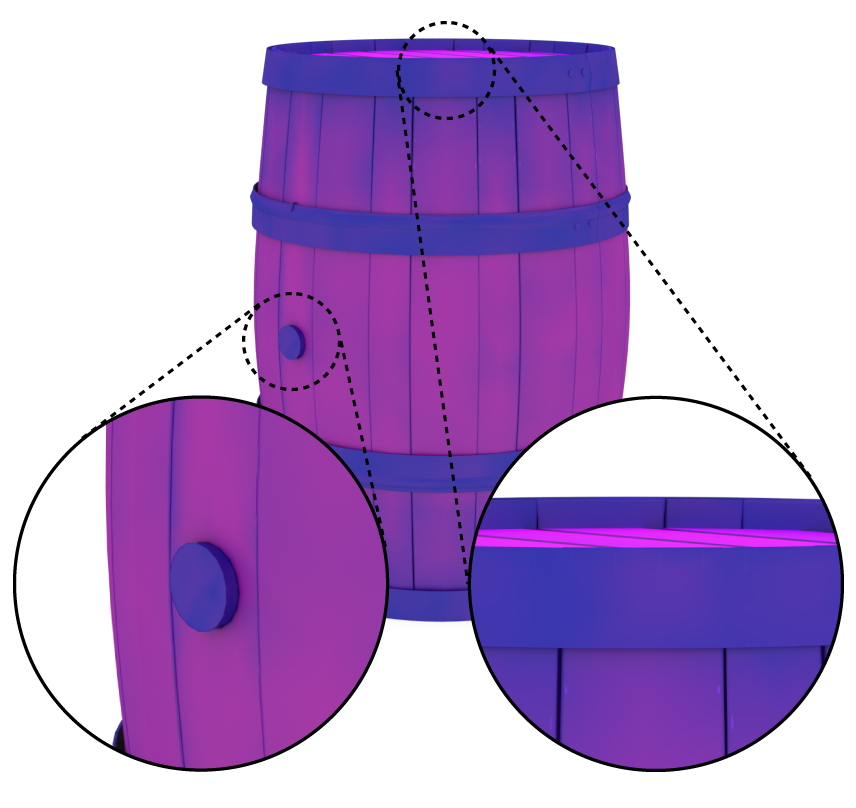}
         \caption{Ours}
         \label{fig:unwrapping_uvs}
     \end{subfigure}
     \begin{subfigure}{0.125\textwidth}
         \centering
         \includegraphics[width=\textwidth]{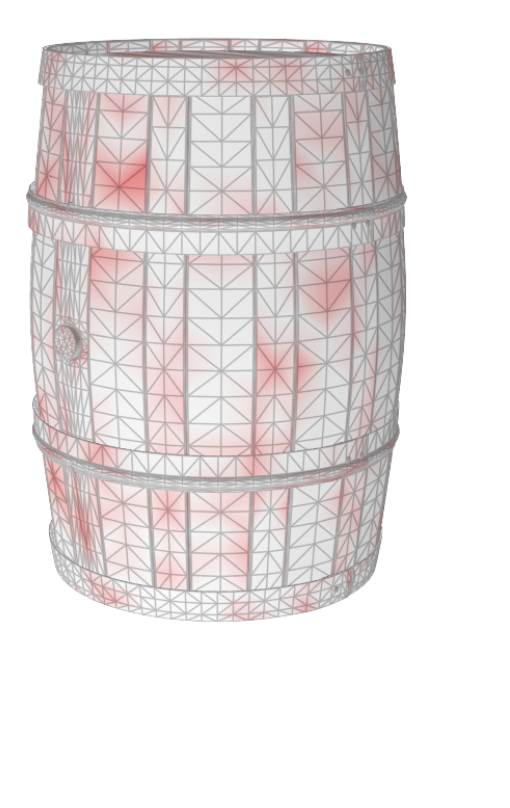}
         \caption{Prediction error}
         \label{fig:unwrapping_uvs}
     \end{subfigure}
    \caption{On this barrel, we compare the precision of predicting the ShDF values $\tilde{\Lambda}$ (b) with the baseline (a). We also show that our approach generates a fairly low error rate compared to the baseline (white: no error, red: error$<0.1\%$).}
    \label{fig:barrel_error}
\end{figure}

\subsection{Performance analysis}
As our approach solely focuses on mesh segmentation, we have limited the analysis to recent methods performing well in that area. We have focused on four datasets to evaluate the precision of our approach: COSEG, Human Body Segmentation (HBS), Autodesk Character Generator, and TurboSquid. We compared the precision of our approach on the segmentation task with PointNet, PointNet++, and MeshCNN.

\begin{table}[h]
    \centering
    \begin{tabular}{|l||c|c|c|c|}
        \hline
        \multirow{2}{*}{\textbf{Method}} & \multicolumn{4}{|c|}{\textbf{Accuracy}}\\\cline{2-5}
          & \textbf{COSEG} & \textbf{HBS} & \textbf{ACG} & \textbf{TS}\\[0.5ex]
        \hline\hline
        \textsl{PointNet} & 91.5\% & 86.5\% & 72.4\% & 58.3\%\\
        \hline
        \textsl{PointNet++} & 94.7\% & 90.8\% & 74.7\% & 61.2\%\\
        \hline
        \textsl{MeshCNN} & 97.3\% & 92.3\% & 78.3\% & 72.8\%\\
        \hline
        \textsl{Ours} & \textbf{97.1\%} & \textbf{94.2\%} & \textbf{92.4\%} & \textbf{86.6\%}\\
        \hline
    \end{tabular}
    \caption{Evaluations comparing our approach with state-of-the-art methods on multiple datasets.}
    \label{table:metrics_STAR}
\end{table}
As presented in Table~\ref{table:metrics_STAR}, our approach performs similarly to MeshCNN and slightly outperforms PointNet and PointNet++ on the COSEG (only the \textsl{Vases} set) and HBS datasets. Also highlighted in that same table, our approach is way more accurate when used on the ACG and TS datasets (even when compared to MeshCNN). This is not surprising as our approach is trained on multiple-resolution samples. By querying the full-resolution neighborhoods and their corresponding downsampled mesh, our approach has proven to be less sensitive to adaptive meshing, which is the case of most of the 3D models contained in the TS dataset.
Lastly, as we aim to speed up the whole traditional mesh segmentation process, we have also compiled a few computation-time results on the presented assets. As shown in Table~\ref{table:performances}, our neural-based approach shows a speed-up factor of up to 10x compared to the original \textit{Shape Diameter Function} method in most of the presented assets.
\begin{table}[t]
    \centering
    \begin{tabular}{|l||c|c|c|c|c|}
        \hline
        \multirow{2}{*}{\textbf{Mesh}} & \multirow{2}{*}{\textbf{Faces}} & \multicolumn{4}{|c|}{\textbf{Accuracy}}\\\cline{3-6}
        &  & \textbf{ShDF} & \textbf{Part.} & \textbf{Ref.} & \textbf{PP}\\[0.5ex]
        \hline\hline
        \textsl{ACM Box} & 1k & 96.5\% & 94.4\% & - & -\\
        \hline
        \textsl{Gas Can} & 1.5k & 95.3\% & 96.9\% & - & -\\
        \hline
        \textsl{Teapot} & 6.3k & 96.9\% & 99.7\% & - & -\\
        \hline
        \textsl{Mayabot} & 56k & 92.4\% & 94.2\% & 95.3\% & -\\
        \hline
        \textsl{Samurai} & 242k & 88.1\% & 89.0\% & 92.4\% & 93.1\%\\
        \hline
        \textsl{Gladiator Hulk} & 1.3M & 87.3\% & 90.7\% & 90.7\% & 91.4\%\\
        \hline
    \end{tabular}
    \caption{Precision metrics obtained on several 3D models. The performances are broken down into four parts: generating ShDF values, partitioning (Part.), refinement (Ref.), and post-processing (PP).}
    \label{table:performance_mesh}
\end{table}
\begin{table*}[h]
    \centering
    \begin{tabular}{|l||c|c|C{1.6cm}|C{1.6cm}|C{1cm}|C{1.6cm}|C{1.6cm}|C{1cm}|}
        \hline
        \multirow{2}{*}{\textbf{Mesh}} & \multirow{2}{*}{\textbf{Vertices}} & \multirow{2}{*}{\textbf{Faces}} & \multicolumn{3}{|c|}{\textbf{Baseline~\cite{shapira2008consistent}}} & \multicolumn{3}{|c|}{\textbf{Ours}}\\\cline{4-9}
        & & & \textbf{ShDF} & \textbf{Partitioning} & \textbf{Total} & \textbf{ShDF} & \textbf{Partitioning} & \textbf{Total}\\[0.5ex]
        \hline\hline
        \textsl{ACM Box} & 0.5k & 1k & 441 & 45 & 486 & \textbf{30} & 26 & \textbf{56}\\
        \hline
        \textsl{Gas Can} & 0.8k & 1.5k & 662 & 46 & 708 & \textbf{44} & 26 & \textbf{70}\\
        \hline
        \textsl{Teapot} & 3.6k & 6.3k & 2836 & 58 & 2894 & \textbf{190} & 33 & \textbf{223}\\
        \hline
        \textsl{Mayabot} & 29.5k & 56k & 23239 & 132 & 23371 & \textbf{1555} & 70 & \textbf{1624}\\
        \hline
        \textsl{Samurai} & 58.1k & 242k & 45689 & 139 & 45828 & \textbf{3057} & 133 & \textbf{3190}\\
        \hline
        \textsl{Gladiator Hulk} & 671k & 1.3M & 528578 & 816 & 529394 & \textbf{35366} & 408 & \textbf{35775}\\
        \hline
    \end{tabular}
    \caption{The timing results are expressed in milliseconds.}
    \label{table:performances}
\end{table*}
Finally, to improve the partitioning part of our approach, we implemented the Gaussian mixture and the clustering to GPU. With our implementation, we almost halved the computation times required by these steps.

\subsection{Dense meshses}
\begin{figure}[h]
    \centering
    \includegraphics[width=\linewidth]{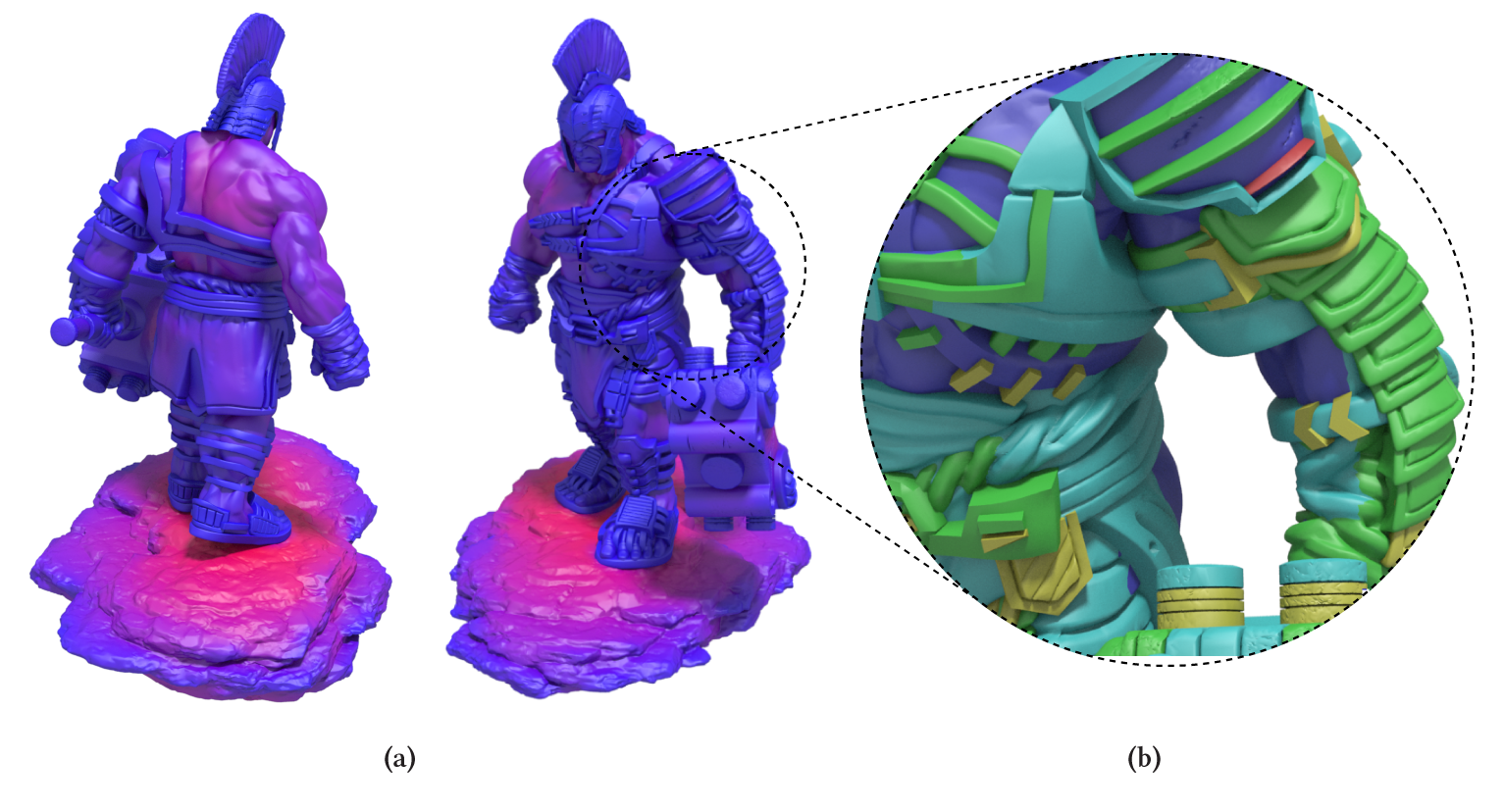}
    \caption{Example of our approach performed on a dense mesh (source: \href{https://desirefx.me/3d_models/gladiator-hulk-mesh-3d-model}{Desire FX}). The generated ShDF values (a) are used to segment the mesh (b).}
    \label{fig:densemesh}
\end{figure}
We also demonstrate that our refinement and post-processing steps can improve the results by a few points on dense meshes. With an error rate of below 2\% compared to the ground truth, we show that our neural ShDF approach performs well on dense meshes. As shown in Table~\ref{table:performance_mesh}, our approach produces high-precision ShDF values on both the \textsl{Samurai} and \textsl{Gladiator Hulk} meshes. We can provide even more precise values using the refinement and post-processing steps. For example, with the \textsl{Gladiator Hulk} mesh (Fig.~\ref{fig:densemesh}), we were able to extract small additional details using the refinement step on segmented parts such as the shoulder pad presented in the close-up (b).

On the \textsl{Samurai} mesh, we obtained better results of the segmentation after post-processing the boundaries (between each segmented part) to smooth them out using alpha expansion based on the dihedral angles between faces (i.e., to measure smoothness and concavity). We decided not to use the refinement and post-processing steps when the resulting segmentation seemed visually adequate.

\subsection{Applications}
\begin{figure}[h]
  \centering
   \includegraphics[width=\linewidth]{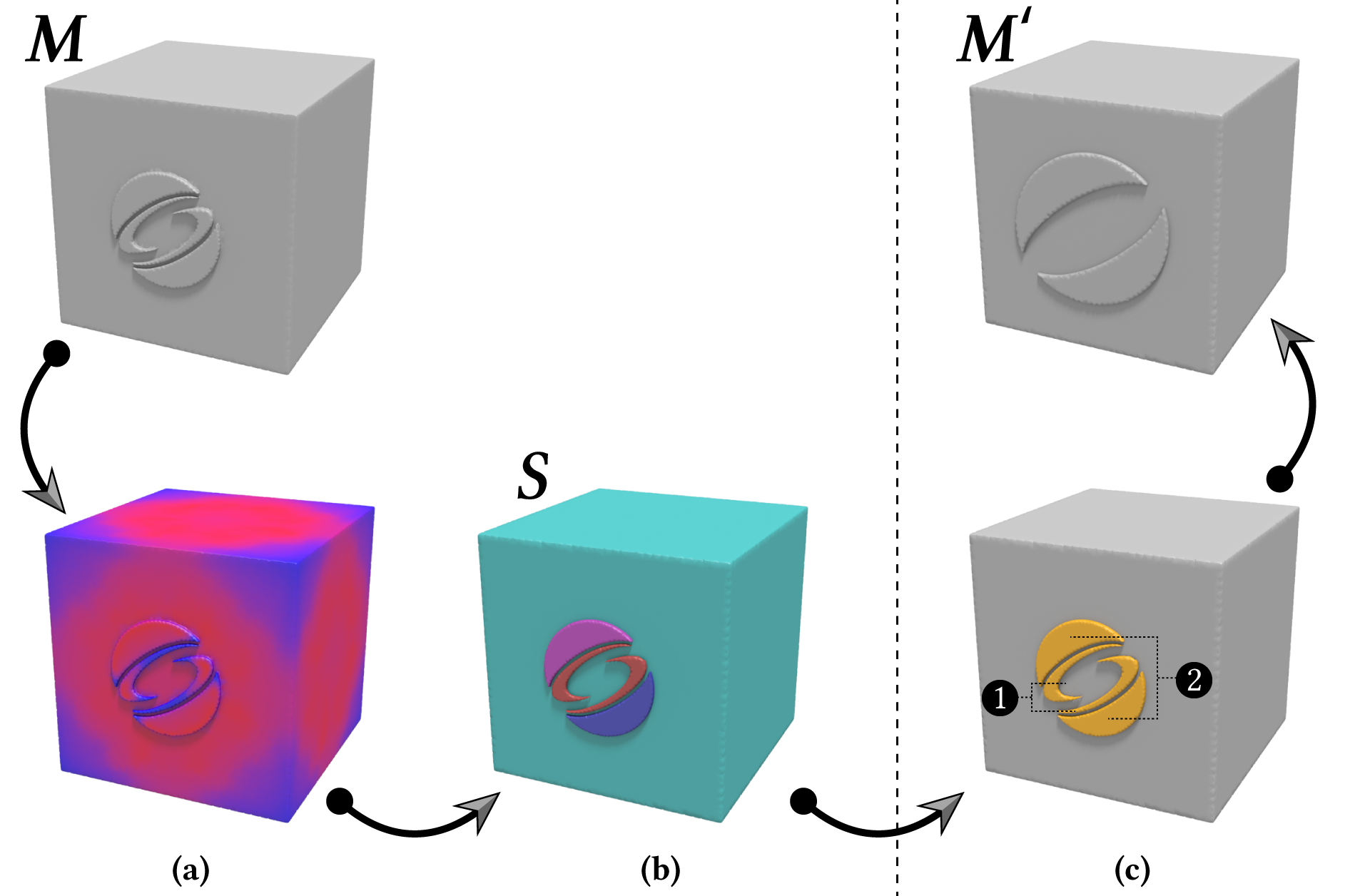}
   \caption{A simple application example of altering an engraved logo in 3ds Max.}
   \label{fig:application_element}
\end{figure}

As previously stated, many tasks intuitively require decomposing meshes before further processing. Mesh segmentation finds many applications in 3D shape analysis such as: texture mapping~\cite{sander2003multi}, 3D shape modeling~\cite{ji2006easy}, 3D shape retrieval~\cite{ferreira2010thesaurus}, multi-resolution and mesh compression~\cite{maglo2011cluster}; and animation~\cite{yuan2016space}. We refer the reader to this survey for further details~\cite{rodrigues2018part} on part-based mesh segmentation. For example, Fig.~\ref{fig:application_element} shows a typical example in which a cube with an engraved logo is decomposed into meaningful parts using our neural ShDF approach (a). Once segmented (b), the parts can then be selected (c) for further manipulations such as: replacing with a different logo, deforming the selected logo, or as shown, scaling up and removing parts of the logo.

\begin{figure}
     \centering
     \begin{subfigure}{0.225\textwidth}
         \centering
         \includegraphics[width=\textwidth]{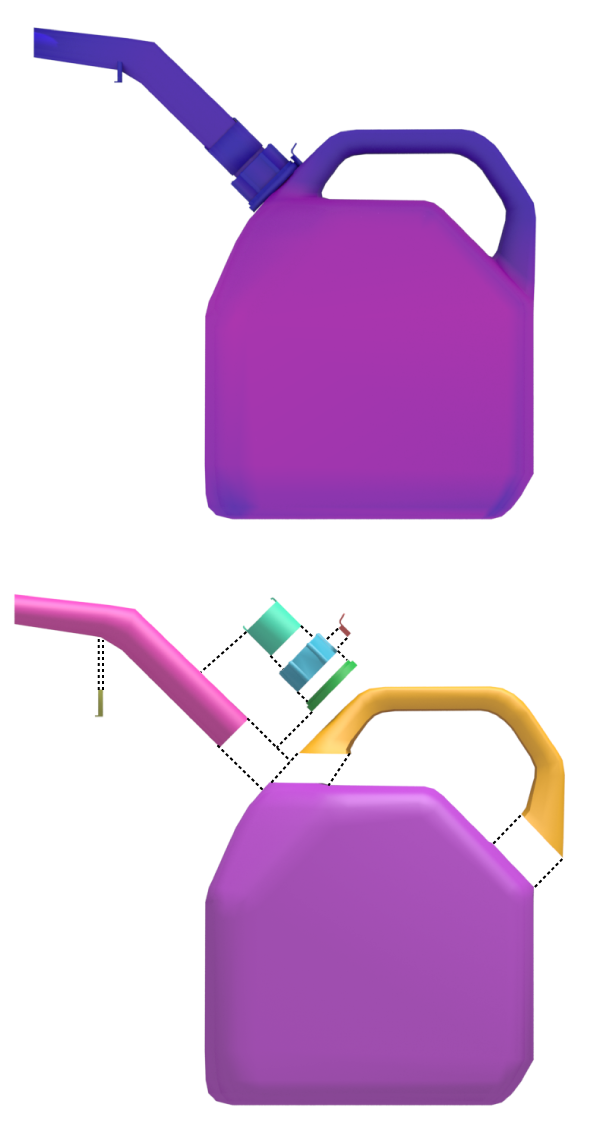}
         \caption{ShDF value (top) and segmented parts (bottom).}
         \label{fig:sdf_seg}
     \end{subfigure}
     \hfill
     \begin{subfigure}{0.225\textwidth}
         \centering
         \includegraphics[width=\textwidth]{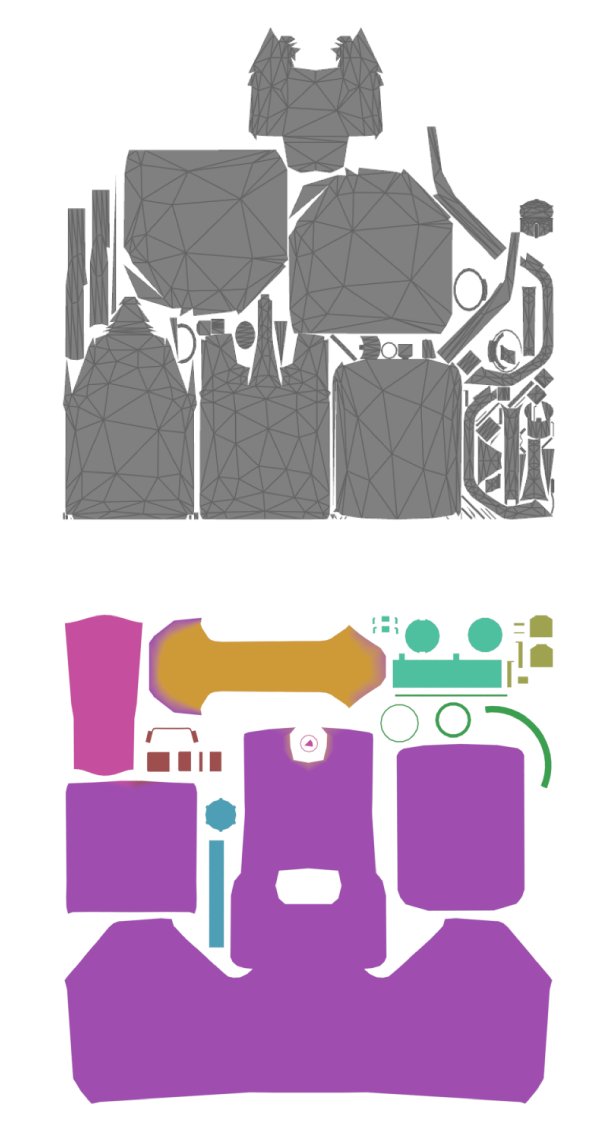}
         \caption{Automatic UV mapping on the whole mesh (top) and recursively on segmented parts (bottom).}
         \label{fig:unwrapping_uvs}
     \end{subfigure}
    \caption{Piece-wise UV unfold.}
    \label{fig:unwrapping}
\end{figure}

We have also experimented to see how effective our approach would be in the case of UV mapping. The unwrapping part is known to be very unintuitive and requires expert knowledge of shape topology to properly unwrap 3D models into a 2D space while preserving the semantic meaning of the parts. Recently, there has been a single attempt to solve this problem by directly predicting UV seams using a graphical attention network (GAT)~\cite{teimury2020graphseam}. Although promising, this proposed method ended up being rather limited to the training set and was hardly generalizable for production uses. We believe that being able to provide semantically meaningful parts on complex models can partially alleviate a fair portion of the pain of the traditional UV mapping workflow. That way, each simpler part can be unwrapped using only a few cuts and simple projection methods. In Fig.~\ref{fig:unwrapping}, we show our approach leveraged to facilitate the 3D unfolding task. As shown, the \textsl{Gas Can} is first segmented into meaningful parts (bottom of Fig.~\ref{fig:sdf_seg}) before running an automatic unwrapping based on the prominent shape features. Our solution from the segmented object (top of Fig.~\ref{fig:unwrapping_uvs}) clearly outperformed the one generated using an automatic method in Maya (bottom image).

\section{Conclusion}
We have presented a \textit{Neural Shape Diameter Function} approach allowing us to considerably accelerate high-quality mesh segmentation while remaining adaptive and controllable. Using the generated ShDF values for our method, we can generate multiple alternative segmentation solutions of the same mesh by adjusting the partitioning parameters for our k-way graph-cut algorithm.

Although there are powerful graph-cut methods implemented on GPU, the use of adaptive tessellation-generated meshes in production to optimize memory makes them less efficient than deep neural networks for segmenting such irregular structures containing a highly variable number of neighbors per vertex (as opposed to images). While our approach partially mitigates this part by querying the mesh at full resolution, we are confident that using stacked networks could still benefit from this approach by leveraging our current network outputs as inputs to a non-binary classifier. We may investigate that architecture as future work.

\bibliographystyle{ACM-Reference-Format}
\bibliography{paper_547}

\appendix

\section{Automating Refinement}
As introduced in Section~\ref{sec:graph_cut}, one of the main advantages of using our approach is that the predicted function for a given mesh can be reused as many times as necessary to produce the desired segmentation.
\begin{figure}[h]
    \centering
    \includegraphics[width=\linewidth]{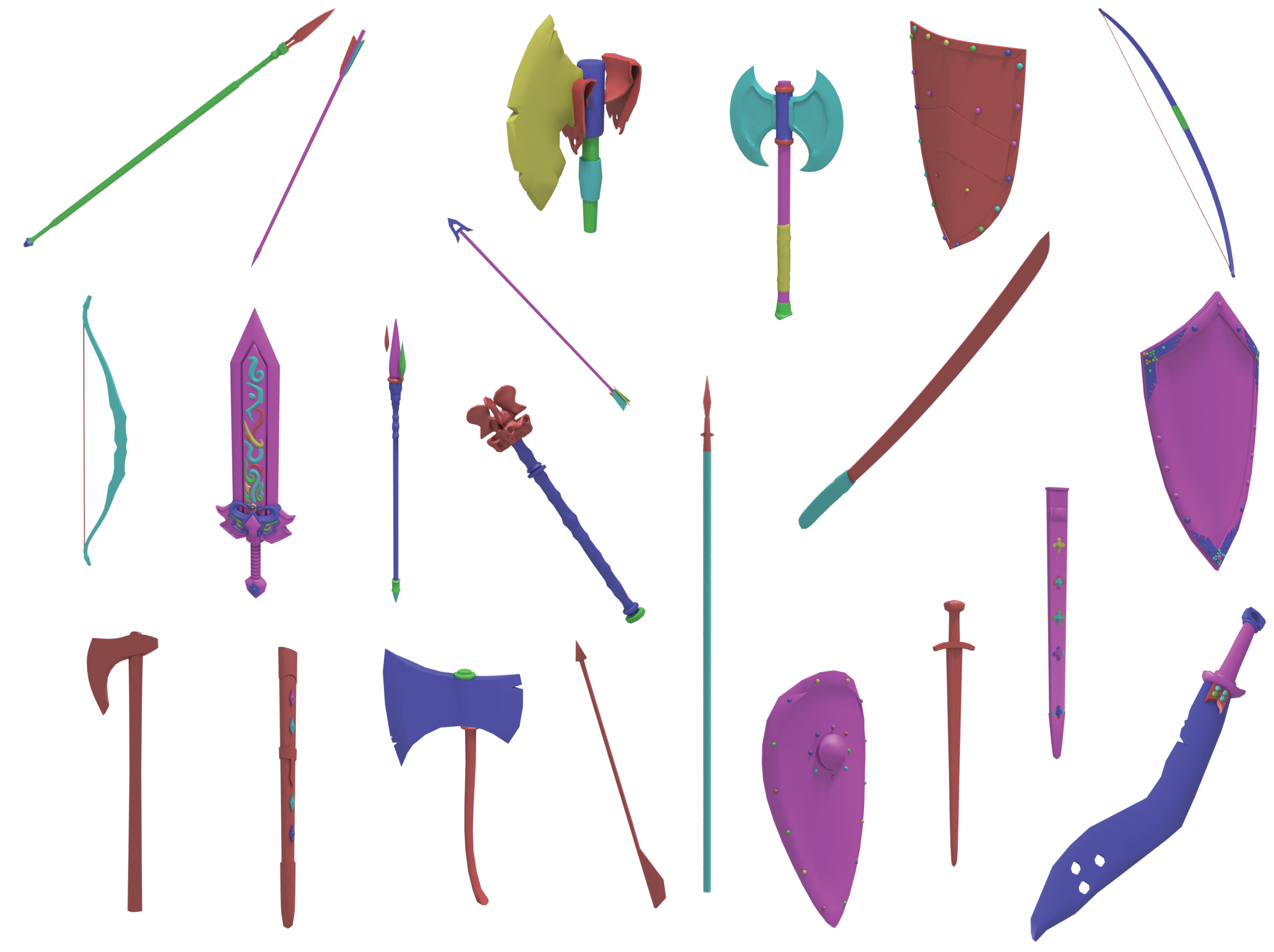}
    \caption{Subset of the TurboSquid (TS) Prop dataset.}
    \label{fig:automation}
\end{figure}

\begin{figure*}[t]
     \centering
     \begin{subfigure}{0.495\textwidth}
         \centering
         \includegraphics[width=\textwidth]{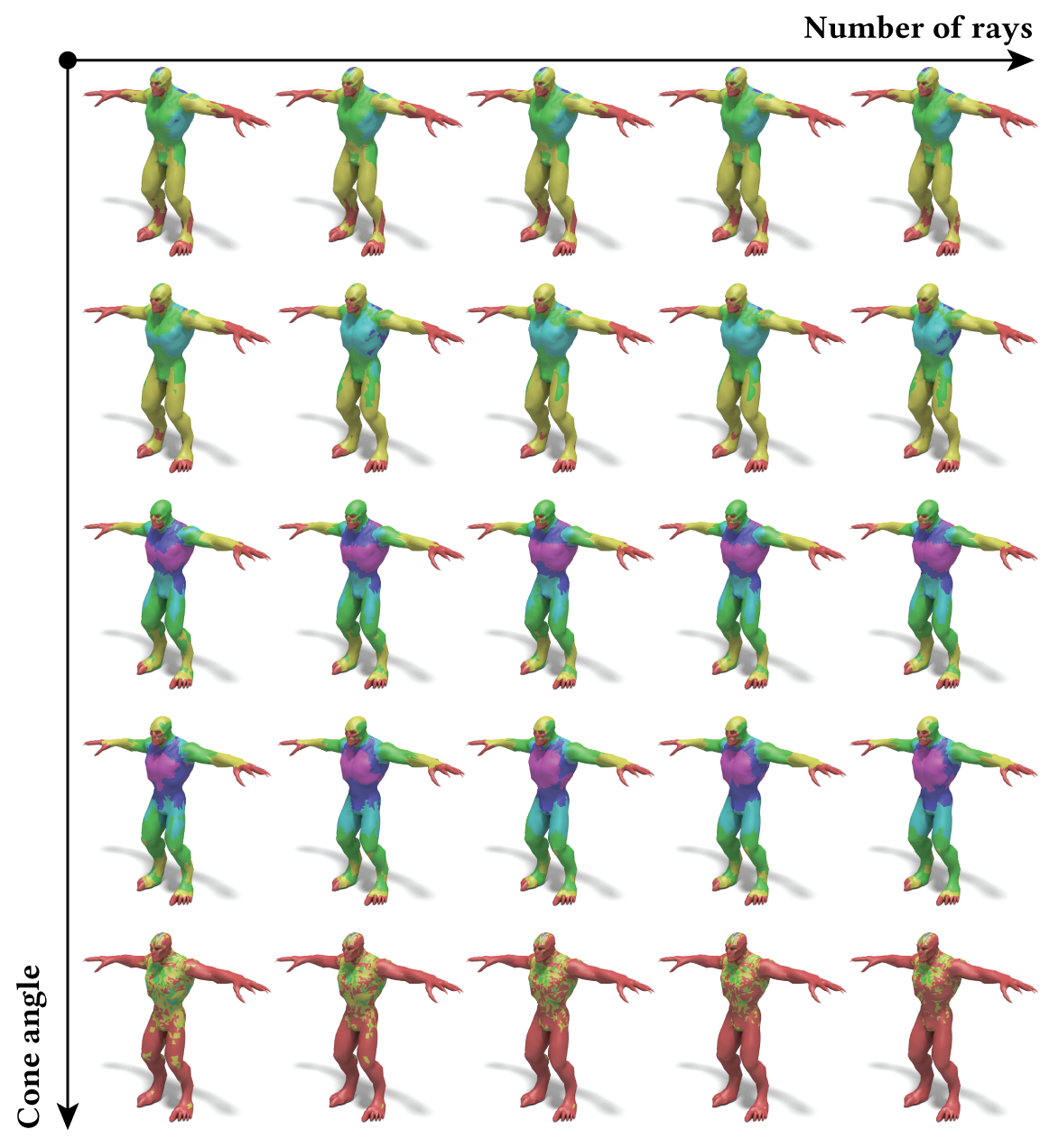}
         \caption{ShDF generation}
         \label{fig:sdf_params}
     \end{subfigure}
     \hfill
     \begin{subfigure}{0.495\textwidth}
         \centering
         \includegraphics[width=\textwidth]{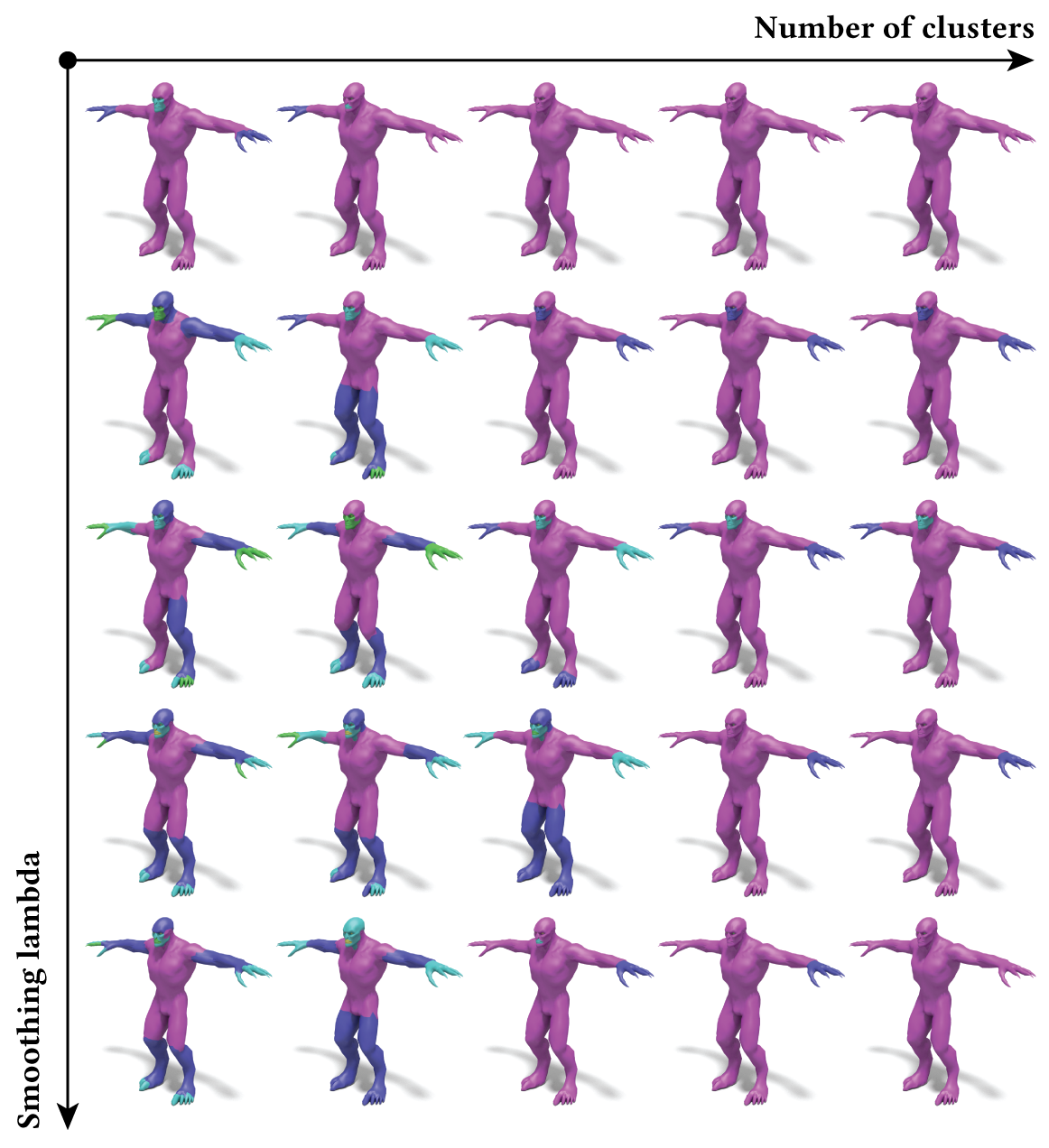}
         \caption{Partitioning}
         \label{fig:seg_params}
     \end{subfigure}
    \caption{Presenting different results on the same base mesh when using different sets of parameters with the baseline approach~\cite{shapira2008consistent}.}
    \label{fig:optimizing_params}
\end{figure*}

The neural ShDF approach considerably reduces the computation times required to find the optimal parameters producing the \textit{shape diameter function} values (Fig.~\ref{fig:sdf_params}). By using this optimal mapping, our approach enables focusing the computational effort to generate the segmentation of the mesh (Fig.~\ref{fig:seg_params}), recursively if necessary.

\begin{figure}[h]
     \begin{subfigure}{0.225\linewidth}
         \centering
         \includegraphics[width=\textwidth]{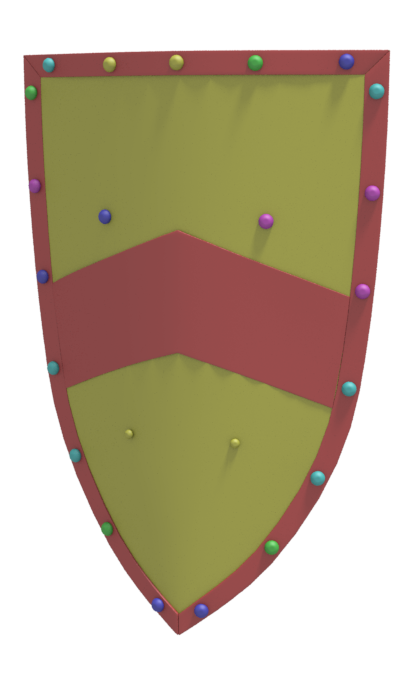}
         \caption{}
         \label{fig:auto_shield}
     \end{subfigure}
     \begin{subfigure}{0.225\linewidth}
         \centering
         \includegraphics[width=\textwidth]{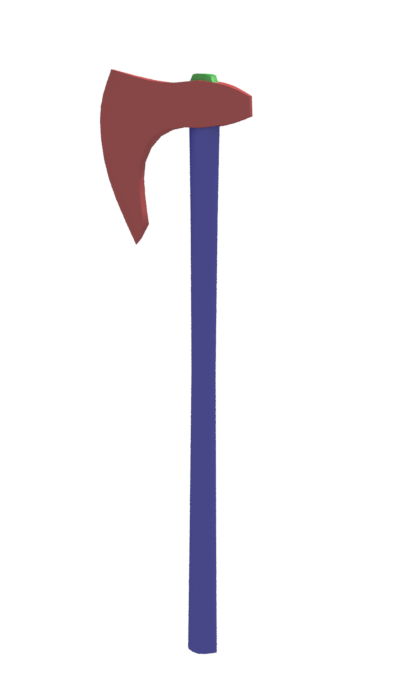}
         \caption{}
         \label{fig:auto_axe}
     \end{subfigure}
    \caption{Refinement step performed on two of the objects from the dataset shown in Fig.~\ref{fig:automation}}
    \label{fig:auto_fixed}
\end{figure}

For example (see Fig.~\ref{fig:automation}), as an experiment on the TS dataset, we have automatically generated the segmentation on a subset of objects using the same partitioning parameters of the k-way graph-cut method. As a result, those parameters were not optimal for a few objects in that subset. Using our approach, we were able to re-generate the segmentation method using suitable parameters on these objects (as shown in Fig.~\ref{fig:auto_fixed}), at almost no additional cost.

\end{document}